\documentclass[12pt]{article}

\usepackage{amssymb}

\newcommand{\lc}{\varepsilon}
\newcommand{\dlr}{\stackrel{\leftrightarrow}{d}{}}
\newcommand{\nablar}{\stackrel{\rightarrow}{\nabla}\!\!{}}
\newcommand{\nablal}{\stackrel{\leftarrow}{\nabla}\!\!{}}
\newcommand{\realni}{\ensuremath{\mathbb{R}}}
\newcommand{\celi}{\ensuremath{\mathbb{Z}}}
\newcommand{\br}{{\realni}}
\newcommand{\ds}{\displaystyle}

\newcommand{\trr}{\triangleright}
\newcommand{\cf}{{\cal F}}
\newcommand{\cg}{{\cal G}}
\newcommand{\cd}{{\cal D}}

\begin{document}

\begin{center}
\textbf{\Large Poincar\'e 2-group and quantum gravity}
\end{center}

\bigskip

\begin{center}
A. MIKOVI\'C\,\footnote{Also at Grupo de Fisica Matem\'atica da Universidade de Lisboa,
Av. Prof. Gama Pinto, 2, 1649-003 Lisboa, Portugal} \\
Departamento de Matem\'atica,
Universidade Lus\'ofona de Humanidades e Tecnologias\\
Av. do Campo Grande, 376, 1749-024 Lisboa, Portugal\\
E-mail: amikovic@ulusofona.pt\\
\end{center}
\centerline{and}
\begin{center}
M. VOJINOVI\'C\\
Grupo de Fisica Matem\'atica da Universidade de Lisboa\\
Av. Prof. Gama Pinto, 2, 1649-003 Lisboa, Portugal\\
E-mail: vmarko@cii.fc.ul.pt
\end{center}

\bigskip
\bigskip

\begin{quotation}
\noindent\small{We show that General Relativity can be formulated as a constrained topological theory for flat 2-connections associated to the Poincar\'e 2-group. Matter can be consistently coupled to gravity in this formulation. We also show that the edge lengths of the spacetime manifold triangulation arise as the basic variables in the path-integral quantization, while the state-sum amplitude is an evaluation of a colored 3-complex, in agreement with the category theory results. A 3-complex amplitude for Euclidean quantum gravity is proposed.}
\end{quotation}

\bigskip
\bigskip
\noindent{\large\bf{1. Introduction}}

\bigskip
\noindent General Relativity (GR) was originally formulated as a dynamical theory of metrics on a spacetime manifold, and it turned out that for a non-perturbative quantization it is more advantageous to reformulate it as a theory of connections, see \cite{Rb} for a comprehensive exposition and references.

More precisely, GR can be represented as a constrained BF theory \cite{cbf}, and this approach led to spin foam formulation of quantum GR, see \cite{Bsfr,Rsf} for reviews and references. The EPRL/FK class of spin foam models \cite{EPRL,FK} allows for a construction of finite quantum gravity transition amplitudes \cite{MVf,FM,H} and the corresponding classical limit is GR \cite{MVea}. However, the absence of the tetrads from the theory makes it difficult to couple massive fermions \cite{Rferm} as well as the gauge fields \cite{Msfym}, so that there is a need for a BF-type reformulation of GR which will include the tetrads.

One way to do this is to introduce the cosmological constant and represent GR as a BF theory for the anti-de Sitter/de Sitter group with a symmetry breaking term, see \cite{Mdsbf} for a review and references. However, the corresponding spin foam perturbation theory is difficult to formulate \cite{MMsfp}, since the symmetry breaking term is the perturbation, and there is no efficient mathematical formalism to calculate the corrections.

Another possible approach is to use the Poincar\'e group, since GR can be represented as a gauge theory for the Poincar\'e group, see \cite{Blag} for a comprehensive exposition and references. However, the transformation law for the tetrads under the local translations does not coincide with the diffeomorphism transformations. Although the diffeomorphism transformations of the tetrads can be represented as suitably restricted local translations, one would like to have a formalism where the local translations coincide with the diffeomorphism transformations of the tetrads.

In this paper we will show that all these problems can be resolved by using the BF theory for 2-groups, also known as BFCG theory, see \cite{GPP,fmm}. The 2-groups are category theory generalizations of the usual groups and the Poincar\'e group can be naturally embedded into a 2-group, see \cite{BH}. We will show that GR can be represented as a constrained BFCG theory for the Poincar\'e 2-group, in analogy with the result that GR is the constrained BF theory for the Lorentz group. Furthermore, the matter fields can be consistently coupled and the path-integral quantization leads to a state sum whose categorical structure agrees with the results derived from the representation theory of the Poincar\'e 2-group \cite{CS,BBFW}.

In section 2 we present the gauge theory of flat connections for the Poincar\'e 2-group, and show that a simple constraint reduce it to GR in the Einstein-Cartan (EC) formulation. In section 3 we show that matter can be coupled such that the GR constraint is preserved. In section 4 we show that the path integral for the topological theory takes a form of a state sum whose amplitude is an evaluation of a colored 3-complex. The 3-complex is associated with the spacetime triangulation, while the colors can be chosen such that they coincide with the irreps, intertwiners and 2-intertwiners for the Euclidean/Poincar\'e 2-group. We then propose a quantum gravity state sum for the Euclidean 2-group case and in section 5 we present our conclusions.

\bigskip
\bigskip
\noindent{\large\bf{2. Poincar\'e 2-group and GR}}

\bigskip
\noindent One way to generalize the notion of a group is to use the category theory, see \cite{BH} for a review and references. A category consists of objects and maps between the objects (called morphisms) such that natural composition rules between the morphisms are satisfied. A 2-category consists of objects, morphisms and maps between morphisms (called 2-morphisms) such that natural composition rules are satisfied. A group is then a category with one object where all morphisms are invertible. Similarly, a two-group is a 2-category with one object where all morphisms are invertible. This abstract definition leads to a concrete realization of a 2-group which is given by a crossed module $(G,H, \partial,\triangleright)$. This is a pair of groups $G$ and $H$, such that $\partial: H \to G$ is a homomorphism and $\triangleright$ is an action of $G$ on $H$ such that certain properties are satisfied, which are direct consequences of the categorical structure, see \cite{BH}. The elements of $G$ represent the 1-morphisms, while the elements of the semidirect product 
$G\times_s H$ represent the 2-morphisms. The canonical example of a 2-group relevant for physics is the Poincar\'e 2-group, where $G = SO(1,3)$, $H={\br}^4$, $\partial$ is a trivial homomorphism and $\triangleright$ is the usual action of the Lorentz transformations on the ${\br}^4$ space. The Lorentz group is the group of morphisms, while the usual Poincar\'e group is the group of 2-morphisms.

One can construct a gauge theory on a 4-manifold $M$ based on a crossed module $(G,H, \partial,\triangleright)$ of Lie groups by using one-forms $A$, which take values in the Lie algebra $\bf g$ of $G$, and 2-forms $\beta$, which take values in the Lie algebra $\bf h$ of $H$ \cite{GPP,fmm}. The forms $A$ and $\beta$ transform under the usual gauge transformations $g: M \to G$ as
\begin{equation}
A \to g^{-1} A g \,+\,g^{-1} d g \,,\quad \beta\to g^{-1} \triangleright \beta \,,\label{tgt}
\end{equation}
while the gauge transformations generated by $H$ are given by
\begin{equation}
A \to A\,+\,\partial\eta\,,\quad\beta\to \beta +d \eta + A \wedge^\triangleright \eta + \eta\wedge \eta \, ,\label{fgt}
\end{equation}
where $\eta$ is a one-form taking values in $\bf h$, see \cite{fmm}. When the group $H$ is Abelian, which happens in the Poincar\'e 2-group case, then the $\eta\wedge\eta$ term in (\ref{fgt}) vanishes, and one obtains the gauge transformations given in \cite{GPP} . 

The pair $(A,\beta)$ represents a 2-connection on a 2-fiber bundle associated to the 2-Lie group $(G,H)$ and the manifold $M$. The corresponding curvature forms are given by
\begin{equation}
{\cal F} = dA + A\wedge A -\partial\beta \,,\quad
{\cal G} = d \beta + A \wedge^\triangleright \beta \,,
\end{equation}
and they transform as
\begin{equation}
{\cal F} \mapsto g^{-1} {\cal F} g\,, \quad {\cal G} \to g^{-1} \trr {\cal G} \,,
\end{equation}
under the usual gauge transformations, while
\begin{equation}
{\cal F} \to {\cal F}\,,\quad {\cal G}  \to {\cal G} +{\cal F} \wedge^\trr \eta\,,
\end{equation}
under the $H$-gauge transformations.

One can introduce a natural topological gauge theory determined by the vanishing of the 2-curvature
\begin{equation}  \cf = 0 \,,\quad \cg =0 \,.
\end{equation}
These equations can be obtained from the action
\begin{equation} 
S_0 =\int_{M} \langle B \wedge \cf \rangle_{\bf g} +  \langle C \wedge \cg \rangle_{\bf h} \,,\label{bfcg}
\end{equation}
where $B$ is a 2-form taking values in $\bf g$, $C$ is a one-form taking values in $\bf h$, $\langle\,,\rangle_{\bf g}$ is a $G$-invariant non-degenerate bilinear form  in $\bf g$ and $\langle\,,\rangle_{\bf h}$ is a $G$-invariant non-degenerate bilinear form  in $\bf h$. The action (\ref{bfcg}) is called BFCG action, in analogy with the BF theory action. The gauge transformations of the Lagrange multiplier fields are given by
\begin{equation}
B \to g^{-1} B g \,,\quad C \mapsto g^{-1} \trr C \,,\label{bct}
\end{equation} 
for the usual gauge transformations, while 
\begin{equation}
B \to B - [C,\eta] \,,\quad C \mapsto C \,,\label{sbcf}
\end{equation}
for the $H$-gauge transformations.

Let us now examine the case of the Poincar\'e 2-group. In this case $A=\omega^{ab}J_{ab}$, $\beta = \beta^a P_a$, where $a,b\in\{0,1,2,3\}$,
$J$ are the generators of the Lorentz group while $P$ are the generators of the translation group ${\br}^4$. Consequently
\begin{equation}
\cf = (d\omega^{ab} + \omega^a{}_c\wedge\omega^{cb} )J_{ab} = R^{ab}J_{ab} \,,\quad \cg = (d\beta^a + \omega^a{}_b \wedge \beta^b)P_a = \nabla\beta^a P_a\,.
\end{equation}
The $G$-gauge transformations are the local Lorentz rotations
\begin{equation}
\omega \to g^{-1} \omega g + g^{-1} dg \,,\quad \beta \to g^{-1} \trr \beta \,,
\end{equation}
while the $H$-gauge transformations are the local translations
\begin{equation}
\delta_\varepsilon \omega =0 \,,\quad \delta_\varepsilon \beta^a = d\varepsilon^a + \omega^a_b \wedge \varepsilon^b \,,
\end{equation}
where $\eta = \varepsilon^a P_a$.

The BFCG action then becomes
\begin{equation}
S_0 = \int_M (B^{ab}\wedge R_{ab} + C_a \wedge \nabla\beta^a )\,, 
\end{equation}
where
\begin{equation}
\delta_\varepsilon B = 0 \,,\quad \delta_\varepsilon C = 0 \,.
\end{equation}
Note that the transformation properties of the one-forms $C^a$ are the same as the transformation properties of the tetrad one forms $e^a$ under the local Lorentz and the diffeomorphism transformations. Hence one can identify the $C$ fields with the tetrads and we write
\begin{equation}
S_0 = \int_M (B^{ab}\wedge R_{ab} + e^a \wedge \nabla \beta_a )\,. \label{tga}
\end{equation}

The action (\ref{tga}) gives a theory of flat metrics, since $R^{ab}=0$ implies the vanishing of the Riemann tensor. In order to obtain GR, we need that only the Ricci tensor vanishes. In the BF theory approach to GR, this problem is resolved by constraining the $B$ field, such that $B^{ab} = \varepsilon^{abcd}e_c \wedge e_d$. Since in the 2-group formulation the tetrads are explicitly present, the required constraint is simply
\begin{equation}
B_{ab} = \varepsilon^{abcd}e_c \wedge e_d \,,
\end{equation}
or
\begin{equation}
B^*_{ab} = e_a \wedge e_b \,.
\end{equation}

Hence the action for GR in the 2-group approach is given by
\begin{equation}
S = \int_M \left(B^{ab}\wedge R_{ab} + e^a \wedge \nabla \beta_a  - \phi_{ab}\wedge (B^{ab} - \varepsilon^{abcd}e_c \wedge e_d)  \right)\,.\label{2pgr}
\end{equation}
The equations of motion are
\begin{equation}
R_{ab} - \phi_{ab} = 0\,, \label{beq}
\end{equation}
\begin{equation}
\nabla\beta_a + 2\lc_{abcd} \phi^{bc}\wedge e^d = 0\,, \label{eeq} 
\end{equation}
\begin{equation}
\nabla B_{ab} - e_{[a} \wedge \beta_{b]} = 0\,, \label{omega}
\end{equation}
\begin{equation} \nabla e_a = 0\,, \label{prvobeta}
\end{equation}
\begin{equation}
B_{ab} - \lc_{abcd} e^c \wedge e^d = 0\,, \label{phi}
\end{equation}
which are obtained by varying $S$ with respect to $B$, $e$, $\omega$, $\beta$ and $\phi$, respectively.

From $B = (e \wedge e)^*$ it follows that $\nabla B \propto (e\wedge \nabla e)^*$, so that $\nabla B =0$ due to (\ref{prvobeta}). The equation (\ref{omega}) then implies that $e_{[a}\wedge \beta_{b]} = 0$. For invertible tetrads we then obtain $\beta =0$, see the Appendix. Therefore (\ref{beq}) and (\ref{eeq}) imply 
\begin{equation}
\varepsilon_{abcd} R^{bc} \wedge e^{d} = 0 \,.\label{ricci}
\end{equation}
The equation (\ref{ricci}) is the same as the equation of motion for the EC action
\begin{equation}
S_{EC} = \int_M \varepsilon_{abcd} e^a \wedge e^b \wedge R^{cd} \,,
\end{equation}
for the $e$ variations, while  (\ref{prvobeta}) is equivalent to $\delta S_{EC}/\delta\omega$ when the tetrads are invertible.

\bigskip
\bigskip
\noindent{\large\bf{3. Coupling of matter}}

\bigskip
\noindent Since the tetrads are present in the BFCG action, the coupling of matter fields is essentially given by the coupling of matter fields in the EC formulation. The only subtlety is in the coupling of fermions, since their presence introduces a non-zero torsion.

The Dirac action for the fermion field in the EC formulation is given by
\begin{equation} \label{DiracDejstvo}
S_D = i \kappa_1 \int \lc_{abcd} \, e^a \wedge e^b \wedge e^c \wedge
\bar{\psi} \left( \gamma^d  \dlr + \{ \omega , \gamma^d \}  
+ \frac{im}{2}\,e^d \right) \psi \, ,
\end{equation}
where $\omega = \omega_{ab} [\gamma^a , \gamma^b]/8$ and $\kappa_1 = 8\pi l_p^2 /3$.  The $\delta (S_{EC} + S_D) / \delta\omega$ equation gives the torsion $T_a \equiv \nabla e_a = - \kappa_2 s_a$, where 
$$
s_a = i\lc_{abcd} \,e^b \wedge e^c\, \bar\psi \gamma_5 \gamma^d \psi \,,
$$
is the spin 2-form, and $\kappa_2 = -3\kappa_1 /4$. Hence in the BFCG formulation we need a term $\int_M \beta_a \wedge s^a$ in the action.

Let us consider the action
\begin{equation} \label{GRDdejstvo}
S_{m} = S + S_D + S_{\beta\psi} \, ,
\end{equation}
where
$$
S_{\beta\psi} =   i\kappa_2 \int \lc_{abcd} e^a \wedge e^b \wedge \beta^c \, \bar{\psi}\gamma_5\gamma^d \psi \, .
$$
By varying $S_m$ with respect to $B$, $e$, $\omega$, $\beta$, $\phi$ and $\bar\psi$, respectively, we obtain
\begin{equation}
R_{ab} - \phi_{ab} = 0 \,,\label{b}
\end{equation}
 \begin{equation}
\nabla\beta_a + \lc_{abcd} e^b \wedge \left[ 2\phi^{cd}  -\frac{3i\kappa_1}{2} \beta^c \bar{\psi}\gamma_5\gamma^d \psi +
  3i\kappa_1 e^c \wedge \bar{\psi} \left( \gamma^d \nablar - \nablal \gamma^d + \frac{im}{6} e^d \right)\psi\right] = 0\,,\label{e}
\end{equation}
 \begin{equation}
\nabla B_{ab} - e_{[a} \wedge \beta_{b]} -2\kappa_2 \lc_{abcd}e^c \wedge s^d = 0\,, \label{o}
\end{equation}
\begin{equation}
\nabla e_a + \kappa_2 s_a = 0\,, \label{drugobeta}
\end{equation}
\begin{equation}
B_{ab} - \lc_{abcd} e^c \wedge e^d = 0\,, \label{fi}
\end{equation}
\begin{equation}
i \kappa_1 \lc_{abcd} e^a\wedge e^b \wedge \left( 2e^c \wedge \gamma^d \nabla + \frac{im}{2} e^c\wedge e^d
 - 3 ( \nabla e^c ) \gamma^d -\frac{3}{4} \beta^c \gamma_5 \gamma^d \right)\psi = 0\,.\label{psi}
\end{equation}

Exactly like in the pure gravity case, from (\ref{fi}) it follows that $\nabla B \propto (e \wedge \nabla e)^*$, so that (\ref{o}) gives
$$
2 \, \lc_{abcd} e^c \wedge \left(\nabla e^d + \kappa_2 s^d \right) + e_{[a} \wedge \beta_{b]} = 0 \,.
$$
This equation gives $e_{[a} \wedge \beta_{b]} =0$, due to (\ref{drugobeta}). If the tetrads are invertible, one then obtains $\beta^a =0$, so that (\ref{e}) gives 
\begin{equation}
\lc^{abcd} e_b\wedge (R_{cd} -T^{\psi}_{cd}) = 0 \,,\label{ede}
\end{equation}
where $T^{\psi}_{ab}$ is the energy-momentum two-form for the fermions. The equation (\ref{ede}) is equivalent to the Einstein equations when a Dirac fermion is coupled to EC gravity.

The  $\delta S_m /\delta\psi$ and $\delta S_m /\delta\bar\psi$ equations are related by the spinor conjugation. For the invertible tetrads, and by using $\nabla e = -\kappa_2 s$, the equation (\ref{psi}) reduces to the usual Dirac equation
\begin{equation} \label{VarijacijaGRDpoPsiovima}
\left(i \gamma^\mu \nabla_{\mu} - m \right)\psi = 0\,, 
\end{equation}
where $\gamma^\mu = e^{\mu}{}_a \gamma^a$.

As far as scalar and YM fields are concerned, they do not couple to $\omega$, so that one simply adds the corresponding EC formalism terms to $S_m$
\begin{equation}
S_m \to S_m + \int_M |e| \left ( g^{\mu\nu} \partial_\mu \Phi \partial_\nu \Phi + g^{\mu\nu}g^{\rho\sigma}\,Tr\,F_{\mu\rho}F_{\nu\sigma}\right)d^4 x \,,
\end{equation}
where $g_{\mu\nu} = e_\mu^a e_\nu^b \eta_{ab}$.

One can also introduce the Immirzi parameter $\gamma$, by adding an additional term $S_{\gamma}$ to the action $S_m$, where
$$
S_{\gamma} = -\frac{1}{\gamma} \int \phi^{ab}\wedge e_a \wedge e_b + \frac{i\kappa_2}{\gamma^2+1} \int \lc_{abcd} e^a \wedge e^b \wedge \beta^c \bar{\psi}\gamma_5 \gamma^d \psi + \frac{i\kappa_2\gamma}{\gamma^2+1} \int e^a \wedge e^b \wedge \beta_a \bar{\psi} \gamma_5 \gamma_b \psi \,.
$$
The resulting equations of motion are equivalent to the equations of motion obtained from the action $S_{EC}+S_D+S_H$, where $S_H$ is the Holst term \cite{Holst1996}
$$
S_H = -\frac{2}{\gamma} \int e^a \wedge e^b \wedge R_{ab} \,.
$$
The physical motivation for the introduction of the Immirzi parameter lies in the fact that it is the coupling constant between fermions and torsion, as discussed in detail in \cite{Freidel2005,Perez2006}.

\bigskip
\bigskip
\noindent{\large{\bf 4. State-sum models}}

\bigskip
\noindent Given the BFCG form of the EC action, one can now proceed to quantize the theory by using the same approach as in the case of spin foam models, see \cite{Rb}. This approach requires first a construction of the state-sum model for the topological theory given by the unconstrained BFCG theory. Then the constraint $B^* = e\wedge e$ has to be imposed on the topological state sum. 

In the topological case one starts from the path-integral
\begin{eqnarray}  Z &=& \int \cd A \,\cd \beta \,\cd B \,\cd C\, \exp\left(i\int_M (\langle B\wedge \cf\rangle  + \langle C\wedge \cg\rangle )\right)\cr
&=& \int \cd A \,\cd \beta \,\delta (\cf) \,\delta (\cg)\,, \end{eqnarray}
see \cite{GPP}. Let $T$ be a regular triangulation of $M$ and $T^*$ the corresponding dual triangulation. Then
\begin{equation}
Z = \int \prod_l dg_l \int \prod_f dh_f \prod_f \delta(g_f) \prod_p \delta(h_p) \,,\label{gintz}
\end{equation}
where $l,f$ and $p$ denote the 1,2 and 3-cells of $T^*$, respectively, and one has
$$
g_l = \exp\left(\int_l A\right)\,,\quad h_f = \exp\left(\int_f \beta\right) \,.
$$

The group elements $g_l \in G$ and $h_f \in H$ represent the corresponding one and 2-holonomies of $A$ and $\beta$, respectively. The group element $g_f = \prod_{l\in \partial f}g_l$ is the holonomy along the boundary of $f$. When the area of $f$ is small, one has
$$
g_f \approx \exp\left(\int_f \cf\right) \,.
$$ 
The group element $h_p$ is the 2-holonomy along the closed surface $\partial p$, and  
$$
h_p = \prod_{f\in\partial p} \tilde h_f \,,
$$
where some of the $\tilde h_f$ are given by $g_l\trr h_f$, where $l\in p$ and $l\notin\partial f$, while the other $\tilde h_f$ are equal to $h_f$, see \cite{GPP}. When the volume of $p$ is small, one has
$$
h_p \approx \exp\left(\int_p \cg \right) \,.
$$

In the case of the Poincar\'e 2-group the integral (\ref{gintz}) can be written as
\begin{equation}
Z = \int \prod_l dg_l \int \prod_f d^4 \vec{x}_f \prod_f \delta(g_f) \prod_p \delta(\vec{x}_p) \,,\label{2pgrint}
\end{equation}
where $\vec x_p = \vec x_f + ... + g_l \vec x_{f'}$ and $f,...,f' \in \partial p$.
The Lorentz group delta function can be expanded by using the Plancherel theorem
$$
\delta(g_f) = \sum_{\Lambda_f} d\mu (\Lambda_f) \,\chi(g_f,\Lambda_f) \,,
$$
where $\Lambda = (j,\rho)$ are the unitary irreducible representations, $\chi$ is the character and $d\mu$ is the appropriate integration measure, see \cite{lgha}. The notation $\delta (g)$ means that the corresponding distribution is concentrated at the identity element of the Lorentz group. The $\delta(\vec x_p)$ is the four-dimensional Dirac delta function and
$$
\delta(\vec x_p) = \frac{1}{(2\pi)^4}\int_{\br^4} d^4 \vec L_p \exp\left(i\vec x_p \cdot\vec L_p \right) \,.
$$

For the sake of simplicity, let us consider the Euclidean case, so that the Poincar\'e 2-group becomes the Euclidean 2-group. The Lorentz group is then replaced by the $SO(4)$ group and $\Lambda =(j^+ , j^-)$ is a pair of $SU(2)$ spins, so that
\begin{equation}
Z = \sum_{\Lambda_f}\int\prod_p d^4 \vec L_p \int \prod_l dg_l \prod_f d^4 \vec x_f \,\dim\,\Lambda_f \,\chi(\Lambda_f,g_f)\,\prod_p e^{i\vec x_p \cdot\vec L_p} \,.
\end{equation}
After integrating $\vec x_f$, we obtain
\begin{equation} 
\begin{array}{ccl}
 Z &=&\ds \sum_{\Lambda_f} \int\prod_p d^4 \vec L_p  \int \prod_l dg_l \,\prod_f \dim\,\Lambda_f \,\chi(\Lambda_f,g_f)\\
&\,&\ds \quad\quad\quad\prod_f \delta\left( g_{l_1(p_1,f)}\vec L_{p_1,f}+g_{l_2(p_2,f)}\vec L_{p_2,f}+g_{l_3(p_3,f)}\vec L_{p_3,f}\right) \,,
\end{array}\label{dss}
\end{equation}
where $p_1,p_2$ and $p_3$ are the three polyhedra which share the face $f$, and $l_1,l_2$ and $l_3$ are the corresponding dual edges satisfying $l_k \in p_k$ and $l_k \notin f$. 

It is instructive to rewrite (\ref{dss}) by using the simplices of $T(M)$
\begin{eqnarray}  Z &=& \sum_{\Lambda_\Delta} \int\prod_\varepsilon d^4 \vec L_\varepsilon  \int \prod_\tau dg_\tau \,\prod_\Delta \dim\,\Lambda_\Delta \,\chi(\Lambda_\Delta,g_\Delta)\cr
&\,& \quad\quad\quad\prod_\Delta \delta\left( g_{\tau_1(\varepsilon_1,\Delta)}\vec L_{\varepsilon_1,\Delta}+g_{\tau_2(\varepsilon_2,f)}\vec L_{\varepsilon_2,\Delta}+g_{\tau_3(\varepsilon_3,\Delta)}\vec L_{\varepsilon_3,\Delta}\right) \,,\label{tss}\end{eqnarray}
where $\varepsilon_k$ are the edges of $\Delta$ and $\varepsilon_k \in \tau_k$ but $\Delta \notin\tau_k$. The delta function 
$$
\delta(g_1 \vec L_1 +g_2 \vec L_2 +g_3 \vec L_3 )
$$
in (\ref{tss}) restricts the integration over the $\vec L$'s whose lengths satisfy the triangle inequalities.
This implies that $L_\varepsilon = |\vec L_\varepsilon|$ can be interpreted as the length of an edge $\varepsilon$.

Note that (\ref{tss}) can be rewritten as
\begin{equation}
Z =  \int\prod_\varepsilon L_\varepsilon^3 dL_\varepsilon \sum_{\Lambda_\Delta, I_\tau} W(L,\Lambda,I)\,,\label{catss}
\end{equation}
where $I_\tau$ is the intertwiner for the four $\Lambda$ of a tetrahedron and
\begin{eqnarray}  \sum_I W(L,\Lambda,I) &=& \int \prod_\varepsilon d\Omega_\varepsilon \int \prod_\tau dg_\tau \prod_\Delta \dim\,\Lambda_\Delta \,\chi(\Lambda_\Delta,g_\Delta)\cr
&\,&\quad\quad\prod_\Delta \delta\left( g_{\tau_1(\varepsilon_1,\Delta)}\vec L_{\varepsilon_1,\Delta}+g_{\tau_2(\varepsilon_2,f)}\vec L_{\varepsilon_2,\Delta}+g_{\tau_3(\varepsilon_3,\Delta)}\vec L_{\varepsilon_3,\Delta}\right)\,. \label{wintegral}\end{eqnarray}
The $d\Omega_\varepsilon$ denotes a 3-sphere volume measure.

The state sums/integrals in (\ref{catss}) will be almost certainly divergent, but what is important is to find the structure of the dual 3-complex amplitude $W$.  The relation (\ref{wintegral}) seems to imply that $W$ will be a function of the $SO(4)$ irreps, but this may not happen because the integral in (\ref{wintegral}) may not be well-defined and a regularization may introduce the irreps for $SO(3)$ and $SO(2)$ subgroups. That this can happen is suggested by the representation theory of the Poincar\'e/Euclidean 2-group on 2-Hilbert spaces, see \cite{CS,BBFW}. Namely, in the Euclidean case the irreps are labeled by positive numbers, which can be identified with the edge lengths $L_\varepsilon$. In the Poincar\'e 2-group case there is also a class of positive-length irreps, and in both cases the corresponding triangle intertwiners are the $SU(2)$ spins when $L_\varepsilon$ form a zero-area triangle. When $L_\varepsilon$ form a non-zero area triangle, then the intertwiners are given by the $U(1)$ spins. The results of \cite{bafr,BW} suggest that the topological amplitude is given by
\begin{equation}
W(L,m,I) = \prod_f A_f (L) \prod_v \frac{\cos S_v(L,m)}{V_v (L)} \,, \label{topw}
\end{equation}
where 
\begin{equation}
S_v(L,m) = \sum_{f;v\in f} m_f \theta_f (L) \,. \label{aregge}
\end{equation} 
The angle $\theta_f (L)$ is the interior dihedral angle for a face $f$ which contains the vertex $v$, $V_v(L)$ is the 4-volume of the four-simplex dual to $v$, $A_f(L)$ is the area of the triangle dual to $f$, $m_f \in \celi$ are the $U(1)$ spins and the 2-intertwiners $I$ are trivial.

Given the topological amplitude (\ref{topw}), one can try to implement the constraint $B^* = e\wedge e$ in order to obtain the state sum for GR, similarly to what was done in the case of spin foam models. Note that $S_v$, given by eq. (\ref{aregge}), has the form of the area-Regge action for a 4-simplex. If we put
\begin{equation}
|m_f|l_P^2 = A_f (L) \,,\label{ssconstr}
\end{equation}
where $l_P$ is the Planck length, then the area-Regge action $S_v(L,m)$ will become the Regge action $S_{vR} (L)$ for a four-simplex. Since the topological vertex amplitude (\ref{topw}) is proportional to $\cos S_v$, which is a sum of $e^{iS_v}$ and $e^{-iS_v}$, then in the state sum amplitude will appear a term proportional to
$$\prod_v e^{iS_v(L)} = \exp \left( i\sum_v S_v (L) \right) = \exp\left( \frac{i}{l_P^2} \sum_f A_f (L) \delta_f (L)\right) \,,$$ 
where $\delta_f$ is the deficit angle and $ S_R = \frac{1}{l_P^2} \sum_f A_f (L) \delta_f (L) $ 
is the Regge action for a manifold triangulation. The appearence of the term in the amplitude proportional to $e^{iS_R}$ is a good sign that the constrained state sum can be further modified such that it corresponds to the path integral for GR.

Therefore we expect that the quantum GR state sum will have a form
\begin{equation}
Z_{GR} = \int\prod_\varepsilon \mu(L_\varepsilon)\, dL_\varepsilon \sum_{m} \prod_\Delta\delta(|m_\Delta|l_P^2 - A_\Delta (L)) \,W_{GR} (L ,m )\,. \label{grss}
\end{equation}
The amplitude $W_{GR}$ and the measure $\mu$ have to be chosen such that $Z_{GR}$ is finite and that the corresponding effective action gives GR in the classical limit. The effective action approach to the semi-classical limit of spin foam models \cite{MVea} suggests that
\begin{equation}
W_{GR}(L,m) = \prod_f A_f (L) \prod_v \frac{e^{iS_v(L,m)}}{V_v (L)} \,,\label{gram}
\end{equation}
i.e., the $\cos S_v$ factor from (\ref{topw}) has to be replaced by $e^{iS_v}$ so that the effective action will have the correct classical limit.

Coupling of matter in the model defined by the state sum (\ref{grss}) will be easier than in the EPRL/FK model case, since the edge lengths $L_\varepsilon$ are explicitly present. One can then use 
$$
W_{\rm matt}(L,m,\varphi) \propto \exp\left(iS_R^{(\rm matt)}(L,\varphi)\right) \,,
$$ 
for the matter amplitudes, where $S_R^{(\rm matt)}$ is the Regge discretized action of a matter field $\varphi$ coupled to gravity. The expressions for $V_\tau (L)$ and $V_\sigma (L)$, which appear in $S_R^{(\rm matt)}$, can be easily obtained, in contrast to the EPRL/FK model case, where the expression for $V_\sigma$ is difficult to write explicitly in terms of the spin foam variables.

As far as the boundary states are concerned, one will have a wavefunction $\Psi(L_\varepsilon , m_\Delta )$ on the boundary $\partial M = \Sigma$, where $\varepsilon,\Delta \in {T(\Sigma)}$ and $T(\Sigma)$ is the triangulation of the three-dimensional manifold $\Sigma$ induced by $T(M)$. By passing to the dual complex $T^*(\Sigma)$, the wavefunction can be written as
$\Psi (L_f, m_l )$, i.e. a function of a colored 2-complex. This reflects the fact that a boundary of a colored 3-complex is a colored 2-complex.

\bigskip
\bigskip

\noindent{\large\bf 5. Conclusions}

\bigskip
\noindent The proposed 2-group reformulation of GR can be used to obtain a category theory generalization of Loop Quantum Gravity. Namely, instead of using the $SU(2)$ spin networks as the basic quantum degrees of freedom, one can use the spin foams with colors $\{L_f, m_l \}$, or more generally, the spin foams with colors $\{L_f, J_l , \iota_v\}$, where $J_l$ could be an $SU(2)$ or a $U(1)$ spin and $\iota_v$ are the corresponding intertwiners. The time evolution will promote a spin foam into a 3-complex with colors $\{L_p, J_f , \iota_l\}$, which we will call the spin cube. The advantage of this generalization is that the edge lengths of a triangulation become the basic dynamical variables. This will facilitate the construction of the path integral such that the classical limit of the corresponding quantum theory is GR and the coupling of matter will be much easier to accomplish. 

The categorical nature of the theory implies that the edge labels of a spacetime triangulation should be the 2-group irreps. The triangle labels should be the corresponding intertwiners and the tetrahedrons should carry the corresponding 2-intertwiner labels. The 2-Hilbert space representation theory for the Poincar\'e/Euclidean 2-group implies that the set of triangle intertwiners $J_\Delta$ should be a set of $SU(2)$ and $U(1)$ spins. The results of \cite{bafr,BW} imply that the Euclidean topological state sum involves only the $U(1)$ spins, while we expect that the Euclidean GR state sum will take the form (\ref{grss}) with the amplitude given by (\ref{gram}). Extension of this result to the Poincar\'e case would give a physical quantum gravity theory.

Note that the representation theory of 2-groups is not unique, since
one can also use the category of chain complexes of vector spaces in order to define the representations, see \cite{fmm,cfm}. The structure of the chain-complex representations is different from the 2-Hilbert space representations, which means that chain-complex representation theory defines an alternative quantization of GR. Hence it would be interesting to develop the chain-complex representation theory of the Poincar\'e/Euclidean 2-group.

The physical significance of 2-Hilbert space representations could be better understood by performing a canonical quantization of the action (\ref{2pgr}). Note that the actions (\ref{2pgr}) and (\ref{GRDdejstvo}) belong to the class of covariant canonical actions for gravity theories introduced in \cite{nest}.

As far as the construction of 4-manifold invariants based on the BFCG state sum is concerned, one would have to regularize the topological state sum/integral based on the amplitude (\ref{topw}) such that the triangulation independence is preserved. One way to do it is to try to implement a gauge-fixing procedure, see \cite{bafr}. Another way is to find a quantum group regularization, since there are strong indications that categorified quantum groups and their representations will be important for the construction of 4-manifold invariants \cite{crfr}. Hence one can try to find a crossed module of Hopf algebras which is a deformation of the Poincar\'e/Euclidean 2-group, and then try to find an appropriate 2-category of representations which will give a finite topological state sum.

\bigskip
\bigskip 
\bigskip
\noindent{\large\bf Acknowledgments}

\bigskip
\noindent We would like to thank J. F. Martins, J. Morton, M. Blagojevi\'c and A. Baratin for discussions. This work has been partially supported by the FCT project PTDC/MAT/099880/2008. MV was also supported by the FCT grant SFRH/BPD/46376/2008.

\bigskip
\bigskip
\noindent{\large\bf Appendix: Proof that $\beta^a=0$}

\bigskip
\noindent We start from the equation
$$
e^{[a} \wedge \beta^{b]}=0 \,.
$$
By working in a coordinate basis, one can rewrite this equation in the component form as
\begin{equation}
\label{KomponenteEwedgeBeta}
\lc^{\mu\nu\rho\sigma} \left( e^a{}_{\mu} \beta^{b}{}_{\rho\sigma}
- e^b{}_{\mu} \beta^a{}_{\rho\sigma} \right) = 0 \, .
\end{equation}

By assuming that $\det(e^a{}_{\mu}) \neq 0$, we denote the inverse tetrads as $e^{\mu}{}_a$. First we contract equation (\ref{KomponenteEwedgeBeta}) with $\lc_{\alpha\nu\gamma\delta} e^{\alpha}{}_a e^{\lambda}{}_b$ to obtain
$$
\lc_{\alpha\nu\gamma\delta} \lc^{\alpha\nu\rho\sigma} \beta^{\lambda}{}_{\rho\sigma} -
\lc_{\alpha\nu\gamma\delta} \lc^{\lambda\nu\rho\sigma} \beta^{\alpha}{}_{\rho\sigma} =0 \, ,
$$
where $\beta^{\lambda}{}_{\rho\sigma} \equiv e^{\lambda}{}_a \beta^a{}_{\rho\sigma}$. Next, we use the identities
$$
\lc^{\lambda\mu\nu\rho} \lc_{\lambda\alpha\beta\gamma} = - \det \left(
\begin{array}{ccc}
\delta^{\mu}_{\alpha} & \delta^{\mu}_{\beta} & \delta^{\mu}_{\gamma} \\
\delta^{\nu}_{\alpha} & \delta^{\nu}_{\beta} & \delta^{\nu}_{\gamma} \\
\delta^{\rho}_{\alpha} & \delta^{\rho}_{\beta} & \delta^{\rho}_{\gamma} \\
\end{array}
 \right) \, , \qquad
\lc^{\mu\nu\rho\sigma} \lc_{\mu\nu\alpha\beta} = -2 \det \left( 
\begin{array}{cc}
\delta^{\rho}_{\alpha} & \delta^{\rho}_{\beta} \\
\delta^{\sigma}_{\alpha} & \delta^{\sigma}_{\beta} \\
\end{array}
 \right) \, ,
$$
in order to eliminate the contractions of the Levi-Civita symbols. After some algebra, we obtain
\begin{equation}
\label{SvedenaJnaZaBeta}
\beta^{\lambda}{}_{\gamma\delta} + \beta^{\sigma}{}_{\gamma\sigma} \delta^{\lambda}_{\delta}
 - \beta^{\sigma}{}_{\delta\sigma} \delta^{\lambda}_{\gamma} = 0\, .
\end{equation}
By contracting the indices $\lambda$ and $\delta$, we immediately obtain $\beta^{\sigma}{}_{\gamma\sigma}=0$. Substituting this back into (\ref{SvedenaJnaZaBeta}), it follows that
$$
\beta^{\lambda}{}_{\gamma\delta} =0\, .
$$
Finally, contracting with the tetrad $e^a{}_{\lambda}$ one obtains the result $\beta^a = 0$.

\end{document}